\documentclass[aps,prd,preprint,showpacs,nofootinbib,eqsecnum,amsmath,amssymb]{revtex4-1}
\usepackage[colorlinks=true,urlcolor=blue,linkcolor=blue,citecolor=blue]{hyperref}
\usepackage{amsmath} 
\usepackage{amssymb}
\usepackage{epsf}
\usepackage{color}
\usepackage{graphicx}
\usepackage{dcolumn}
\usepackage{bm}

\begin{document}

\title{Search for manifestations of spin-torsion coupling}

\author{\firstname{Mariya Iv.}~\surname{Trukhanova}}
\email{trukhanova@physics.msu.ru}
\affiliation{M. V. Lomonosov Moscow State University, Faculty of Physics, Leninskie Gory,  Moscow, Russia}
\affiliation{Theoretical Physics Laboratory, Nuclear Safety Institute,
Russian Academy of Sciences, B. Tulskaya 52, 115191 Moscow, Russia}

\author{Pavel Andreev}
\email{andreevpa@physics.msu.ru}
\affiliation{M. V. Lomonosov Moscow State University, Faculty of Physics, Leninskie Gory,  Moscow, Russia}

\author{\firstname{Yuri N.}~\surname{Obukhov}}
\email{obukhov@ibrae.ac.ru}
\affiliation{Theoretical Physics Laboratory, Nuclear Safety Institute,
Russian Academy of Sciences, B. Tulskaya 52, 115191 Moscow, Russia}

\begin{abstract}
We investigate the axial vector torsion-spin coupling effects in the framework of the Poincar\'e gauge theory of gravity with the general Yang-Mills type Lagrangian. The dynamical equations for the ``electric'' and ``magnetic'' components of the torsion field variable are obtained in the general form and it is shown that the helicity density and the spin density of the electromagnetic field appear as the physical sources. The modified Maxwell's equations for the electromagnetic field are derived, and the electromagnetic wave propagation under the action of the uniform homogeneous torsion field is considered. We demonstrate the Faraday effect of rotation of the polarization for such a wave, and establish the strong bound on the possible cosmic axial torsion field from the astrophysical data.
\end{abstract}

\maketitle

\section{Introduction}

The search for new spin-dependent interactions between fundamental particles apart from the magnetic dipole interaction is an important area of high-energy physics research beyond the Standard Model \cite{Safronova}. \'E. Cartan was the first who proposed, at the beginning of the 20th century, the post-Riemannian geometrical structures generated by the microstructural properties of the physical matter, particularly to analyze the coupling of the torsion of spacetime to the intrinsic spin \cite{Cartan}. 

The interest in the theory of gravitation with spin and torsion based on the Riemann--Cartan geometry had considerably grown in the second half of the 20th century after the consistent gauge theory formalism was developed \cite{2,21,3,4,41}. It is now well established that the spacetime torsion can only be detected with the help of the spin \cite{YS,HOP,OP}. The early theoretical analysis of the possible experimental manifestations of the torsion field can be found in \cite{Hehl,9,Poincare}. The so-called Einstein--Cartan theory \cite{FWH1,FWH2,Tra} with the linear Hilbert--Einstein gravitational Lagrangian represents a degenerate version of the Poincar\'e gauge gravity. In this model, the torsion couples to spin algebraically and, therefore, it vanishes outside the matter sources, but essentially modifies the physical structure inside astrophysical compact objects, see \cite{Ray1,Ray2,Ray3}.

The torsion becomes a dynamical propagating field in the Poincar\'e gauge gravity theory with a Yang--Mills type Lagrangian \cite{AA}, and the most general gravitational model with the Lagrangian (which is quadratic in the curvature and torsion) was considered in Reference \cite{Poincare2} with an emphasis on the consistency of the gauge theory of gravity with experimental observations at the macroscopic level. Accordingly, its probing should be essentially confined to the microscopic level and focus on the study of the dynamics of fundamental particles, atoms, and molecules. It is worthwhile to note that it has not yet been possible to create a source of spin density that could generate torsion to be detected in the laboratory. However, one can establish the constraints on the spin--torsion coupling, in particular from the experimental search for the Lorentz and $CPT$ violation. The bounds on new spin-dependent interactions were found \cite{Heckel,Heckel1} with the help of a torsion pendulum technique, which was also used in the search for $CP$-violating interactions between the pendulum's electrons and unpolarized matter in laboratory surroundings or the Sun. Among other physical effects, the contribution of the interacting vector and pseudovector of the torsion to the hyperfine splitting of the ground state of the hydrogen atom was evaluated in \cite{Poincare3,Poincare4}, whereas a possible manifestation of the spin--torsion coupling in the scattering of polarized photons in a medium of sodium vapor was analyzed in \cite{Poincare5}. The experimental upper bounds on the spin--torsion interactions were reported in \cite{Lehnert,Lamm,ostor}. 

In the context of the growing interest in fundamental physics at the sub-eV scale, Moody and Wilczek \cite{Moody} analyzed new fields, generating dipole couplings between fermions that can be detectable in laboratory experiments, paying special attention to axions. An axion  as a hypothetical particle was postulated in the Peccei--Quinn theory to resolve the strong $CP$ problem in quantum chromodynamics, and it can produce long-range dipole forces \cite{Peccei}. Similar dipole interactions between fermions can be produced by other particles \cite{10,11,12,13}, for example, by an arion, which is a boson corresponding to a spontaneous breaking of the chiral lepton symmetry \cite{13}. The search for axions and axion-like particles is of considerable interest in relation to the cold dark matter issue. Currently, there is a number of experimental attempts to find axions that encompass the Primakoff effect for the astrophysical axions \cite{Primakoff}, the polarization measurements for light propagating in a magnetic field, the light shining through wall experiments \cite{Wall}, and the cosmic axion spin precession experiment \cite{Axionspin,Axionspin2}.

Another direction in the search for long-range spin-dependent interactions is the prediction of the existence of an unparticle in the context of quantum excitations of scale-invariant interactions \cite{Unparticle}, along with exotic spin-1 bosons or paraphotons \cite{Paraphoton}, which are currently being actively investigated \cite{Paraphoton1}. Quite generally, the analysis of the behaviors of atomic systems affected by new hypothetical spin-dependent forces gives rise to the constraints on the coupling constants \cite{Safronova} with the sixteen types of potentials characterizing interactions between fermions mediated by the new exotic particles \cite{Paraphoton1,Dobrescu}. A wide range of relevant laboratory investigations was carried out using the physical methods of atomic and molecular systems, as well as make use of the optical methods: ion capture experiments \cite{ioncapture}, using nitrogen-vacancy centers in diamonds \cite{diamond,diamond2}, based on molecular and atomic spectroscopies \cite{spectroscopy}. 

The gauge symmetry is one of the most powerful physical concepts underlying the description of fundamental interactions. The global gauge $U(1)$ invariance in quantum electrodynamics leads to the electric charge conservation law, whereas the local gauge invariance requires the introduction of a massless vector field that mediates the long-range interaction between charges, with the electric current as a source of the electromagnetic field. In quantum chromodynamics, the invariance under the global group $SU(2)$ yields the isospin conservation law, and the invariance under the local transformations of the group $SU(2)$ leads to the introduction of the  Yang--Mills gauge field. The extension of this theory leads to the explanation of the strong interactions in terms of the exchange of gluons.

Recently \cite{Naik,Malik}, an attempt was made to construct a gauge theory model to describe the weak spin--spin interactions. It was suggested \cite{Naik} that the invariance of the Lagrangian under the local Lorentz transformation requires the introduction of a massless axial vector gauge field, which gives rise to a super-weak long-range spin--spin force in a vacuum, which is attractive for parallel spins. In this model, the axial vector field couples to the axial vector current of the Dirac fermion field and the photon field or a neutral spin-1 field. The axial vector field was introduced \cite{Pradhan} to provide stability of the classical electron and to construct divergence-free quantum electrodynamics. Optical experiments are the most accurate and accessible for measuring the effects of new physical fields. The direct interactions between electromagnetic and gauge fields were considered in \cite{Yakushin} in the Poincar\'e gauge gravity approach, whereas the coupling of the photon to the axial vector gauge field was supported by the local Lorentz symmetry group in the approach \cite{Naik,Malik}. Here, we study the possible influence of the axial torsion field on the propagation and polarization of an electromagnetic wave.

The structure of the paper is as follows. In Section~\ref{General theory}, we briefly outline the corresponding Lagrange--Noether framework of the Poincar\'e gauge gravity theory. In Section~\ref{El wave}, we study the propagation of the electromagnetic wave under the influence of the uniform homogeneous axial vector torsion field. Finally, in Section~\ref{conc}, we discuss the results obtained and apply them to derive the strong upper limit of the value of the cosmic background torsion field from the astrophysical data. In Appendix~\ref{appA}, the structure of the general quadratic Poincar\'e gauge gravitational field Lagrangian is given, and the effective coupling constants are introduced. 

Our basic conventions and notations are as follows. The world indices are labeled by Latin letters $i,j,k,\ldots = 0,1,2,3$ (for example, the local spacetime coordinates $x^i$), whereas we reserve Greek letters for tetrad indices, $\alpha,\beta,\ldots = 0,1,2,3$ (i.e., for labeling the legs of an anholonomic coframe $e^\alpha_i$). In order to distinguish separate tetrad indices, we put hats over them. Finally, spatial indices are denoted by Latin letters from the beginning of the alphabet, $a,b,c,\ldots = 1,2,3$. The metric of the Minkowski spacetime reads $g_{ij} = {\rm diag}(c^2, -1, -1, -1)$, and the totally antisymmetric Levi--Civita tensor $\eta_{ijkl}$ has the only nontrivial component $\eta_{0123} = c$, so that $\eta_{0abc} = c\varepsilon_{abc}$ with the three-dimensional Levi--Civita tensor $\varepsilon_{abc}$. The spatial components of the tensor objects are raised and lowered with the help of the Euclidean three-dimensional metric $\delta_{ab}$.

\section{Gauge theory of gravitation} \label{General theory}

The Poincar\'e gauge gravity \cite{PRs,Blagojevic:2002,Blagojevic:2013,Obukhov:2006,Obukhov:2018,primer} is an extension of Einstein's general relativity theory (GR), in which the spin, energy, and momentum are independent sources of the gravitational fields (the metric $g_{ij}$ and connection $\Gamma_{ki}{}^j$), and the spacetime structure is described by the Riemann--Cartan geometry with the curvature and the torsion:
\begin{align}
R_{kli}{}^j &= \partial_k\Gamma_{li}{}^j - \partial_l\Gamma_{ki}{}^j + \Gamma_{kn}{}^j \Gamma_{li}{}^n - \Gamma_{ln}{}^j\Gamma_{ki}{}^n,\label{curv}\\
T_{kl}{}^i &= \Gamma_{kl}{}^i - \Gamma_{lk}{}^i.\label{tors}
\end{align}
The Riemann--Cartan connection can be decomposed into the Riemannian and the post-Riemannian parts, 
\begin{equation}
\Gamma_{kj}{}^i = \widetilde{\Gamma}_{kj}{}^i - K_{kj}{}^i,\label{dist}
\end{equation}
where the Christoffel symbols are determined by the metric
\begin{equation}\label{Chr}
\widetilde{\Gamma}_{kj}{}^i = {\frac 12}g^{il}(\partial_jg_{kl} + \partial_kg_{lj} - \partial_lg_{kj}),
\end{equation}
and the contortion tensor is constructed in terms of the torsion
\begin{equation}\label{NTQ}
K_{kj}{}^i = -\,{\frac 12}(T_{kj}{}^i + T^i{}_{kj} + T^i{}_{jk})\,.
\end{equation}
The torsion (\ref{tors}) can be decomposed \cite{primer} into three irreducible components,
\begin{equation}
T_{kl}{}^{i} = {}^{(1)}T_{kl}{}^{i} + {}^{(2)}T_{kl}{}^{i} + {}^{(3)}T_{kl}{}^{i},
\end{equation}
where the second irreducible part features the torsion trace vector
\begin{equation}
{}^{(2)}T_{kl}{}^{i} = {\frac 13}\!\left(\delta^i_{k}T_{l} - \delta^i_{l}T_{k}\right),\label{tT2}
\end{equation}
the third irreducible part is constructed in terms of the torsion axial pseudo-vector
\begin{equation}
{}^{(3)}T_{kl}{}^{i} = -\,{\frac 13}\,\eta_{kl}{}^{ij}\overline{T}{}_j,\label{tT3}
\end{equation}
and the first irreducible purely tensor part has the properties
\begin{equation}
{}^{(1)}T_{ik}{}^{i} = 0,\qquad {}^{(1)}T_{ijk}\eta^{ijkl} = 0.\label{tT1}
\end{equation}
The vector and pseudovector of torsion are defined as
\begin{equation}
T_j := T_{ij}{}^i,\qquad \overline{T}{}^j = {\frac 12}T_{kli}\eta^{klij}.\label{Pv}
\end{equation}

Here, we focus on the dynamical realization of the Poincar\'e gauge theory as a Yang--Mills type model with the most general quadratic in curvature and torsion Lagrangian (\ref{LRT}), see Appendix~\ref{appA} for the details. Earlier \cite{Yakushin,Poincare4}, the contributions of the vector and the pseudovector (\ref{Pv}) to physical effects at the microscopic level were analyzed in the framework of this theory and the strong constraints were established on the spin--torsion coupling parameters. Following \cite{Yakushin}, we continue to study the influence of the {\it axial pseudovector} torsion field ${}^{(3)}T_{kl}{}^i$ on physical matters, and assume that the metric of spacetime is flat, whereas possible post-Riemannian deviations of the spacetime geometry are small. 

As a result, the connection (\ref{dist}) reduces to the contortion, $\Gamma_{kj}{}^i = - K_{kj}{}^i = {\frac 12}{}^{(3)}T_{kj}{}^i$, and by combining (\ref{tT3}) and (\ref{curv}), we find (for the curvature) that $R_{kli}{}^j = {\frac 13}\eta_i{}^{jn}{}_{[k}\partial_{l]}\overline{T}_n$, for the small post-Riemannian corrections. Then it is straightforward to verify that the Yang--Mills-type gauge gravity Lagrangian (\ref{LRT}) is simplified to
\begin{equation}
L = \hbar\left\{-\,{\frac 14}f_{ij}f^{ij} + {\frac {\mu^2}2}\,\alpha_i\alpha^i - {\frac \lambda 2}
(\partial_i\alpha^i)^2\right\}.\label{LRTa}
\end{equation}
Here, $f_{ij} = \partial_i\alpha_j - \partial_j\alpha_i$ is constructed from the rescaled axial torsion trace vector field
\begin{equation}
\alpha_i = {\frac {\ell_\rho}{3}}\sqrt{\frac{-\Lambda_5}{2\kappa c\hbar}}\,\overline{T}_i,
\end{equation}
and the coupling constants (\ref{L12})-(\ref{mu3}) of the Poincar\'e gravity Lagrangian (\ref{LRT}) determine 
\begin{equation}\label{mula}
\mu^2 = -\,{\frac{3\mu_1}{\ell_\rho^2\Lambda_5}},\qquad \lambda = {\frac {3\Lambda_4}{2\Lambda_5}}.
\end{equation}

It is known that the particle spectrum of the Yang--Mills type Poincar\'e gauge gravity model (\ref{LRT}) contains, in general, the so-called ghost and tachyon modes that may lead to the loss of stability and unitarity of the theory. These issues were analyzed in References \cite{Vasilev,Beltran,Belyaev}, and the necessary stability conditions were derived that restrict the choices of the coupling constants. Accordingly, we here specialize to the class of models with $b_3 = 2b_1$, which for the spin 1 sector yields $\lambda = 0$, thus avoiding the stability and unitarity problems. Furthermore, we assume that $a_2 = a_1 - a_0$, which corresponds to the vanishing rest mass $\mu = 0$, in agreement with the estimates derived for the axial vector field \cite{Belyaev,Mohanty} from the high-energy physics phenomenology.

\subsection{Interaction between fermions and axial torsion}

In accordance with the minimal coupling principle \cite{ostor}, the interaction between the fermion field $\psi$ and gauge fields is introduced via the spinor covariant derivative $D_i\psi$, where
\begin{equation}\label{Ds}
D_i = \partial_i - {\frac {iq}{\hbar}}A_i + \frac{i}{4}\Gamma_i{}^{\alpha\beta}\sigma_{\alpha\beta},
\end{equation}
with the Lorentz group generators $\sigma^{\alpha\beta} = i\gamma^{[\alpha}\gamma^{\beta]}$ constructed from the Dirac matrices $\gamma^\alpha$. As a result, the dynamics of the spinor field coupled with the gauge fields on the Minkowski flat metric background is given by the Lagrangian
\begin{equation}\label{Dirac_field}
L_D = \frac{i\hbar}{2}\left\{\overline{\psi}\gamma^i\partial_i\psi - (\partial_i\overline{\psi})\gamma^i\psi
\right\} - mc\overline{\psi}\psi + qA_i\overline{\psi}\gamma^i\psi
+ {\frac 34}\hbar\chi\alpha_i\overline{\psi}\gamma^i\gamma_5\psi.
\end{equation}
Here, we used the identity $\gamma^{\mu}\sigma^{\alpha\beta} + \sigma^{\alpha\beta}\gamma^{\mu} = -\,2\epsilon^{\mu\nu\alpha\beta}\gamma_{\nu}\gamma_{5}$ and introduced 
\begin{equation}\label{chi}
\chi = {\frac{1}{\ell_\rho}}\sqrt{\frac {2\kappa c\hbar}{-\Lambda_5}}.
\end{equation}
Thereby, the axial pseudovector torsion field naturally couples to the spinor axial current $j^i_f=\overline{\psi}\gamma^i\gamma_5\psi$ or to the spin and helicity of the Dirac fermion. By recalling the definition of the Planck length $\ell_{\rm Pl}$, we can recast (\ref{chi}) into
\begin{equation}\label{chi1}
\chi = {\frac{\ell_{\rm Pl}}{\ell_\rho}}\sqrt{\frac {16\pi}{-\Lambda_5}}.
\end{equation}
which demonstrates that the spin--torsion coupling constant $\chi$ is {\it very small}, provided we assume that the characteristic length of the Poincar\'e gauge gravity is much larger than the Planck scale, $\ell_\rho\gg\ell_{\rm Pl}$.

\subsection{Interaction between the electromagnetic field and axial torsion}

Following Pradhan et al. \cite{Naik,Malik,Pradhan}, the interaction of the axial torsion and the electromagnetic field is derived from the standard Maxwell--Lorentz Lagrangian when the ordinary derivatives are replaced by covariant ones. An apparent breaking of the $U(1)$ gauge invariance can be fixed by the modified Stueckelberg method \cite{Stueckelberg}. Together with the dynamical Lagrangian for the axial vector field (\ref{LRTa}) and the fermion sector terms (\ref{Dirac_field}), the total Lagrangian for the torsion field interacting with the spin of the matter sources then reads \cite{Malik}
\begin{align}
L =&\, -\,\frac{1}{4}\sqrt{\frac {\varepsilon_0}{\mu_0}}\,F_{ij}F^{ij} + \frac{i\hbar}{2}\left\{
\overline{\psi}\gamma^i\partial_i\psi - (\partial_i\overline{\psi})\gamma^i\psi\right\}
- mc\overline{\psi}\psi + qA_i\overline{\psi}\gamma^i\psi \nonumber\\
&+\,\hbar\Bigl\{-\,{\frac 14}f_{ij}f^{ij} + {\frac {\mu^2}2}\,\alpha^2 - {\frac \lambda 2}
(\partial\alpha)^2\Bigr\} + {\frac 34}\hbar\chi\alpha_i\overline{\psi}\gamma^i\gamma_5\psi +
\chi\sqrt{\frac {\varepsilon_0}{\mu_0}}\,\alpha_i\eta^{ijkl}A_j\partial_kA_l,\label{Lagrangian1}
\end{align}
where $\varepsilon_0$ and $\mu_0$ are the electric and magnetic constants of the vacuum. The axial vector torsion is thereby coupled to the axial current density of the electron and photon fields:
\begin{equation}\label{current}
j^i_f = \overline{\psi}\gamma^i\gamma_5\psi,\qquad j^i_b = \eta^{ijkl}A_j\partial_kA_l,
\end{equation}
where we explicitly have
\begin{equation}\label{AA}
A_i = \{-\,\phi, \bm{A}\}, \qquad \alpha_i = \{-\,\varphi, \bm{\alpha}\}.
\end{equation}
The quantization of the model (\ref{Lagrangian1}) was analyzed in \cite{Yakushin} and the static potential between fermions (due to the exchange of the axial torsion) was computed.

\subsection{The electromagnetic source of an axial vector field}

Let us consider the spin densities of the spinor and electromagnetic fields as the sources of the axial vector field $\alpha_i$. The components of the axial currents (\ref{current}) for fermions and photon fields can be obtained by substituting Dirac bispinors $\psi = \biggl(\begin{array}{c}u \\ v\end{array}\biggr)$ into Equation (\ref{current}). 
\begin{equation}
j^i_f = -\left\{(u^*v+v^*u)/c,\  (u^*\bm{\sigma}u + v^*\bm{\sigma}v)\right\}, \qquad
j^i_b = {\frac 1c}\left\{\bm{A}\cdot\bm{B},\  (\bm{E}\times\bm{A} + \phi\bm{B})\right\}.
\end{equation}

\subsection{Field equations}

Neglecting the fermion sector, let us derive the field equations for the Lagrangian (\ref{Lagrangian1}). Technically, we need to make variations with respect to the axial torsion field $\alpha_i$ and the electromagnetic field potential $A_i$. The corresponding Euler--Lagrange equation $\delta L/\delta\alpha_i = 0$ for the torsion reads:
\begin{equation}\label{eqT}
-\,\partial_jf^{ij} + \mu^2\alpha^i + \lambda\partial^i(\partial\alpha)
+ {\frac {\chi}{\hbar}}\sqrt{\frac {\varepsilon_0}{\mu_0}}\,\eta^{ijkl}A_j\partial_kA_l = 0.
\end{equation}
Introducing the ``electric'' and ``magnetic'' components of the torsion variable by means of
\begin{equation}\label{EBt}
{\mathcal E}_a = f_{a0},\qquad {\mathcal B}^a = {\frac 12}\epsilon^{abc}f_{bc},\qquad a,b = 1,2,3, 
\end{equation}
along with the usual definitions of the electric and magnetic fields
\begin{equation}\label{EBm}
E_a = F_{a0},\qquad B^a = {\frac 12}\epsilon^{abc}F_{bc},\qquad a,b = 1,2,3, 
\end{equation}
we recast (\ref{eqT}) into the three-dimensional form:
\begin{align}
\bm{\nabla}\cdot\bm{\mathcal E} + \mu^2\varphi - \lambda\partial_t(\partial\alpha) &= 
-\,{\frac{\chi}{\hbar\mu_0}}\,\bm{A}\cdot\bm{B},\label{eqT1}\\
\bm{\nabla}\times\bm{\mathcal B} -{\frac 1{c^2}}\partial_t\bm{\mathcal E} + \mu^2\bm{\alpha}
+ \lambda\bm{\nabla}(\partial\alpha) &= -\,{\frac {\chi\varepsilon_0}{\hbar}}\left(\phi\bm{B} 
+ \bm{E}\times\bm{A}\right).\label{eqT2}
\end{align}

In a similar way, we derived the modified Maxwell equations as the Euler--Lagrange equation $\delta L/\delta A_i = 0$:
\begin{equation}\label{eqM}
-\,\partial_jF^{ij} + {\frac{\chi}{2}}\,\eta^{ijkl}A_jf_{kl} - \chi\,\eta^{ijkl}\alpha_jF_{kl} = 0.
\end{equation}

An immediate observation is in order. Since $\partial_i\partial_jF^{ij} = 0$ identically (symmetric lower indices contracted with the antisymmetric upper indices), by taking the divergence $\partial_i$ of the field equations (\ref{eqT}) (in the special class of stable and unitary models under consideration with $\lambda = 0$ and $\mu^2 = 0$) and (\ref{eqM}), we derive
\begin{equation}\label{FF0}
\eta^{ijkl}F_{ij}F_{kl} = 0,\qquad \eta^{ijkl}F_{ij}f_{kl} = 0,
\end{equation}
respectively. Making use of (\ref{EBt}) and (\ref{EBm}), we find that only crossed-field configurations are actually allowed:
\begin{equation}\label{EB0}
\bm{E}\cdot\bm{B} = 0,\qquad \bm{E}\cdot\bm{\mathcal B} + \bm{\mathcal E}\cdot\bm{B} = 0.
\end{equation}
In particular, this includes wave configurations. 

By making use of (\ref{AA}), (\ref{EBt}), and (\ref{EBm}), we rewrite the inhomogeneous Maxwell equations (\ref{eqM}) in the three-dimensional form:
\begin{align}
\bm{\nabla}\cdot\bm{E} &= 2\chi c\,\bm{\alpha}\cdot\bm{B}
- \chi c\,\bm{A}\cdot\bm{\mathcal B},\label{eqM1}\\
\bm{\nabla}\times\bm{B} - {\frac 1{c^2}}\partial_t\bm{E} &= 
{\frac{2\chi}{c}}\left(\varphi\bm{B} + \bm{E}\times\bm{\alpha}\right) -
{\frac{\chi}{c}}\left(\phi\bm{\mathcal B} + \bm{\mathcal E}\times\bm{A}\right).\label{eqM2}
\end{align}
As usual, we have to add the homogeneous Maxwell system,
\begin{align}
\bm{\nabla}\cdot\bm{B} &= 0,\label{eqMh1}\\
\bm{\nabla}\times\bm{E} + \partial_t\bm{B} &= 0.\label{eqMh2}
\end{align}

\section{Influence of axial torsion on electromagnetic wave}\label{El wave}

Here, we focus on the analysis of the dynamics of the electromagnetic field under the action of the background axial torsion, whereas the full coupled system will be considered elsewhere. Among the possible background configurations, of special interest are the cases of the uniform homogeneous field and the wave configurations. The uniform background field may arise on macroscopic scales, mimicking a distinguished cosmic frame violating the Lorentz symmetry, similar to the mechanisms discussed in \cite{VAK1,VAK2,Carroll,Itin}. It seems natural to turn to the case of the uniform axial torsion background that was extensively considered in earlier literature \cite{Lamm,Belyaev,Mohanty}

\subsection{The case of the uniform external axial torsion field}

Assuming the uniform axial torsion field, when the components $\alpha_i = \{-\,\varphi, \bm{\alpha}\}$ are constant in time and do not change in space, we can solve Maxwell's equations for the plane wave ansatz
\begin{equation}\label{wave}
\bm{E} = \bm{E}_0e^{-i\omega t + i\bm{k}\cdot\bm{r}},\qquad \bm{B} = \bm{B}_0e^{-i\omega t + i\bm{k}\cdot\bm{r}}.
\end{equation}
Substituting this into the homogeneous system, (\ref{eqMh1}) and (\ref{eqMh2}), we derive $\bm{k}\cdot\bm{B} = 0$, and
\begin{equation}\label{BkE}
\bm{B} = {\frac {\bm{k}\times\bm{E}}{\omega}}, 
\end{equation}
and making use of this in (\ref{eqM2}), we obtain the algebraic equation
\begin{equation}\label{eqME}
\bm{k}\times(\bm{k}\times\bm{E}) + {\frac {\omega^2}{c^2}}\,\bm{E}
+ i\,{\frac{2\chi}{c}}\left(\varphi\bm{k} - \omega\bm{\alpha}\right)\times\bm{E} = 0.
\end{equation}
Evaluating the determinant, we find the dispersion relation
\begin{equation}\label{disp1}
\left({\frac {\omega^2}{c^2}} - k^2\right)^2{\frac {\omega^2}{c^2}} -
\left({\frac {\omega^2}{c^2}} - k^2\right)u^2 - (ku)^2 = 0.
\end{equation}
Here we denoted
\begin{equation}
\bm{u} := {\frac{2\chi}{c}}\left(\varphi\bm{k} - \omega\bm{\alpha}\right),
\end{equation}
whereas $k^2 = \bm{k}\cdot\bm{k}$, $u^2 = \bm{u}\cdot\bm{u}$, and $(ku) = \bm{k}\cdot\bm{u}$. It is worthwhile to notice that, similar to the wave in a vacuum, the magnetic field is orthogonal to the electric field and the wave vector,
\begin{equation}\label{BE0}
\bm{B}\cdot\bm{E} = 0,\qquad \bm{B}\cdot\bm{k} = 0,
\end{equation}
which follows from (\ref{BkE}). However, the electric field is not orthogonal to the wave vector, in general,
\begin{equation}\label{kE}
i\bm{k}\cdot\bm{E} = 2\chi c\,\bm{\alpha}\cdot\bm{B}
= {\frac {2\chi c}{\omega}}\,\bm{\alpha}\cdot(\bm{k}\times\bm{E}),
\end{equation}
which is derived by substituting the plane wave ansatz (\ref{wave}) into the inhomogeneous equation (\ref{eqM1}); this is also a direct consequence of (\ref{eqME}). 

{\bf Special case 1}. Assuming $\varphi \neq 0, \bm{\alpha} = 0$, from (\ref{disp1}), we find a simpler dispersion relation
\begin{equation}\label{disp2}
\left({\frac {\omega^2}{c^2}} - k^2\right)^2 - \left({\frac{2\chi}{c}}\right)^2\varphi^2k^2 = 0.
\end{equation}
This can be immediately recast into 
\begin{equation}\label{disp2a}
{\frac {\omega^2}{c^2}} = \bm{k}\cdot\bm{k} \pm {\frac{2\chi}{c}}\,
\varphi\,\sqrt{\bm{k}\cdot\bm{k}}.
\end{equation}
The second term on the right-hand side describes a deformation of the light cone under the influence of the torsion component $\varphi$.

{\bf Special case 2}. Assuming $\varphi = 0, \bm{\alpha} \neq 0$, the general result (\ref{disp1}) reduces to
\begin{equation}\label{disp3}
\left({\frac {\omega^2}{c^2}} - k^2\right)^2 - \left(2\chi\right)^2
\left[\left({\frac {\omega^2}{c^2}} - k^2\right)\alpha^2 + (k\alpha)^2\right] = 0.
\end{equation}
Here, $\alpha^2 = \bm{\alpha}\cdot\bm{\alpha}$, and $(k\alpha) = \bm{k}\cdot\bm{\alpha}$. The resulting dispersion relation can be straightforwardly simplified into
\begin{equation}\label{disp3a}
{\frac {\omega^2}{c^2}} = \bm{k}\cdot\bm{k} + 2\chi^2\,\bm{\alpha}\cdot\bm{\alpha}
\pm 2\chi\sqrt{\chi^2(\bm{\alpha}\cdot\bm{\alpha})^2
+ (\bm{k}\cdot\bm{\alpha})^2} = 0.
\end{equation}

\subsection{Rotation of the polarization plane}

Without loss of generality, one can assume that the electromagnetic wave propagates along the $z$-axis, in other words, we take $\bm{k} = (0, 0, k_z)$. It is more convenient to analyze the two special cases separately. In the first case $(\varphi \neq 0, \bm{\alpha} = 0)$, the dispersion relation (\ref{disp2a}) yields two values for the wave vector,
\begin{equation}\label{k1}
k_z = k_{\pm} = {\frac {\omega \pm \chi\varphi}{c}},
\end{equation}
in the leading order of the small coupling constant $\chi$. Hence, there are two independent waves propagating along $z$ with two different phase velocities. As a result, after extracting the corresponding amplitudes $\bm{E}_\pm$ for the electric field from the algebraic equation (\ref{eqME}), we find the solution 
\begin{equation}\label{solE1}
\bm{E} = E_+e^{-i\omega t + ik_+z}\,\bm{e}_+ + E_-e^{-i\omega t + ik_-z}\,\bm{e}_-\,,
\end{equation}
where we denote the combinations of the basis vectors
\begin{equation}\label{epm}
\bm{e}_\pm = {\frac {\bm{e}_x \mp i\bm{e}_y}{2}}.
\end{equation}
Therefore, from the point of view of physics, the solution (\ref{solE1}) describes the superposition of the right-hand (counter-clockwise) circularly polarized and the left-hand (clockwise) circularly polarized waves. 

\begin{figure}[h]
\centering
\includegraphics[width=0.75\textwidth]{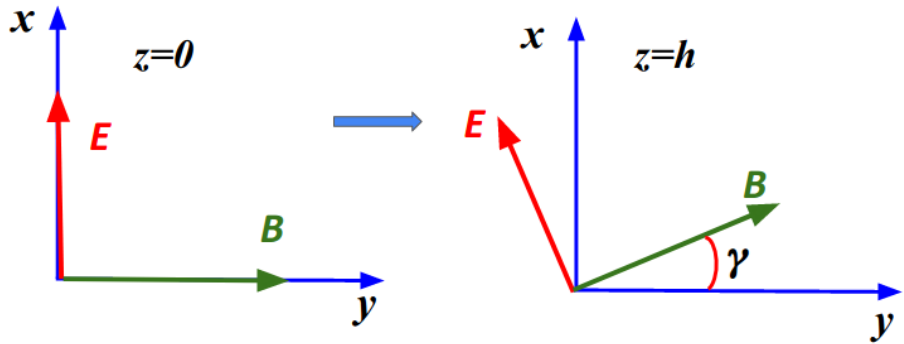}
\caption{Faraday effect of polarization rotation under the action of uniform axial torsion.}\label{Faraday}
\end{figure}

Recalling that the linearly polarized wave arises as the sum of the right-hand and left-hand circular waves with equal amplitudes, $E_+ = E_-$, we then obtain the real solution
\begin{equation}\label{solEl1}
\bm{E} = E_0\,\cos[\omega(t - z/c)]\left\{\cos(\chi\varphi z/c)\,\bm{e}_x -
\sin(\chi\varphi z/c)\,\bm{e}_y\right\}.
\end{equation}
Thus, we recover the {\it Faraday effect} when the polarization vector continuously rotates with the propagation of the plane wave. The polarization rotation angle $\gamma$, see Figure~\ref{Faraday}, from the initial point $z=0$ until the point $z=h$ is determined by 
\begin{equation}\label{angle1}
\gamma = {\frac{\chi\varphi h}{c}}.
\end{equation}

For the second case of the axial torsion field configuration $(\varphi = 0, \bm{\alpha} \neq 0)$, the form of the solution depends on the relative orientation of the $\bm{\alpha}$ with respect to the wave vector of the electromagnetic wave $\bm{k}$. Quite generally, given the direction of the wave propagation, we can decompose $\bm{\alpha} = \bm{\alpha}_{||} + \bm{\alpha}_\bot$ into the longitudinal and transversal projections on the wave vector $\bm{k}$ and the plane orthogonal to it. Accordingly, we derive the solution for the real electric field of the linearly polarized electromagnetic plane wave, to the first order in the interaction constant $\chi$:  
\begin{align}
\bm{E} =&\, E_0\,\cos[\omega(t - z/c)]\left\{\cos(\chi\alpha_{||}z)\,\bm{e}_x - \sin(\chi\alpha_{||}z)
\,\bm{e}_y\right\} \nonumber\\
&-\,{\frac {2\chi\alpha_{\bot}c}{\omega}}\,E_0\,\sin[\omega(t - z/c)]
\sin(\chi\alpha_{||}z + \phi_0)\,\bm{e}_z\,.\label{solEl2}
\end{align}
Since the wave propagates along the $z$-axis, we have $\bm{k}\cdot\bm{\alpha} = k_z\alpha_{||}$, and
\begin{equation}\label{alpha}
\alpha_{||} = \bm{e}_z\cdot\bm{\alpha} = \alpha\cos\theta,\qquad \alpha_{\bot} = \alpha\sin\theta,
\end{equation}
and $\cos\phi_0 = \bm{e}_x\cdot\bm{\alpha}_{\bot}$ measures the angle between the projection $\bm{\alpha}_{\bot}$ and the basis vector $\bm{e}_x$. Obviously, by choosing the coordinate frame appropriately, we can always make $\phi_0 = 0$. It is worthwhile to note that, in accordance with (\ref{kE}), the electric field has a nontrivial component along the wave vector and the third term in (\ref{solEl2}) vanishes only when the wave propagates along the axial torsion $\bm{\alpha}$. In that case, the polarization rotation angle
\begin{equation}\label{angle2}
\gamma = \chi\alpha_{||}h = \chi\alpha\cos\theta h  
\end{equation}
from the initial point $z=0$ to the endpoint $z=h$ is maximal. However, when the axial vector field is orthogonal to the direction of propagation of the electromagnetic wave $\bm{\alpha}\perp\bm{k}$, there will be no rotation of the polarization plane. This generalized Faraday effect is supported by the solution of the dispersion equation (\ref{disp3}),
\begin{equation}\label{k2}
k_z = k_{\pm} = {\frac {\omega} c} \pm \chi\alpha_{||} = {\frac {\omega} c} \pm \chi\alpha\cos\theta,
\end{equation}
in the leading order of the small coupling constant $\chi$, which gives rise to the two waves traveling with two different phase velocities.

\section{Discussion and conclusions}\label{conc}

The search for exotic new forces and interactions generated by the spin of matter particles, fields, and continuous media has a long history. Since the corresponding spin--torsion coupling constant $\chi$ is very small, the detection of such new forces becomes a challenging issue and requires high-precision measurements. On the other hand, the new fields should be truly highly penetrating.

Taking into account that optical experiments are among the most accurate ones, the analysis of possible optical effects of the axial vector torsion field appears to be quite promising. Here, the classical dynamics of the axial vector field were studied in the framework of the Yang--Mills type Poincar\'e gravity model with a focus on the interaction with the electromagnetic field. We derived the dynamical equations for the axial vector torsion field, (\ref{eqT1}) and (\ref{eqT2}), and identified the source of the ``electric'' component of the torsion variable with the helicity density of the electromagnetic field $\sim\bm{A}\cdot\bm{B}$, which characterizes non-trivial topological properties of the field configuration, whereas the ``magnetic'' component of the torsion variable is generated by the spin density of the electromagnetic field. 

Continuing the earlier studies of the spin--torsion effects in the fermion sector \cite{Poincare3,Poincare4,Poincare5}, we here turn to the boson sector. Maxwell's equations are modified (\ref{eqM1})-(\ref{eqMh2}) in the presence of the axial torsion field. The analysis of the propagation of electromagnetic waves under the action of the axial vector torsion field reveals the Faraday effect of the rotation of the wave's polarization and the angle of rotation is determined by the coupling constant, the magnitude of the axial vector torsion field and the travel distance: $\gamma = {\frac{\chi\varphi h}{c}}$ or $\gamma = \chi\alpha_{||}h$. This is consistent with the similar effect arising due to the Lorentz symmetry violation \cite{VAK1,VAK2} or due to the action of the pseudoscalar axion \cite{Carroll,Itin}.

Compared to earlier literature that focused on the evaluation of the spin--spin interaction potential \cite{AA,Moody,12,Paraphoton1,Dobrescu,Yakushin,Naik,Malik}, the results obtained provide a qualitatively new approach to the search for possible manifestations of the spin--torsion coupling with the help of the optical polarization methods. When discussing the most general model of the interacting electromagnetic field and propagating axial torsion field \cite{Fabbri:2015,Fabbri:2021}, one has to pay special attention to the acausal (i.e., superluminal propagation) anomalies. In agreement with the conclusions of \cite{Fabbri:2015,Fabbri:2021}, the class of models under consideration with $b_3 = 2b_1$ yields $\lambda = 0$ for the axial pseudovector field, and so the stability, unitarity, and causality issues are safely avoided. 

Assuming the cosmic nature of the background axial torsion, we can apply the results obtained in the astrophysical situation, and analyze the distribution over the sky sphere of the polarization of radiation, coming from distant radio galaxies. Then taking the observational data collected in Table I of Reference~\cite{Carroll}, and repeating verbatim the computations by replacing the Lorentz-violating parameter with the axial torsion pseudovector, we establish the strong bounds on the torsion's magnitude: $|\overline{T}| \lesssim 8.7 \times 10^{-27}\,$m$^{-1}$. This turns out to be significantly lower than the limits found from the analysis of the spin--torsion coupling effects in the fermion sector \cite{9,Poincare3,Poincare4,Poincare5,Lehnert,Lamm,ostor}. 

\begin{acknowledgments}
We express our gratitude to I.~A. Kudryashov (Nuclear Physics Research Institute of Moscow State University) for discussing the problem. The work of MIT was supported by the Russian Science Foundation under grant 22-72-00036.
\end{acknowledgments}


\appendix
\section{Poincar\'e gauge gravity dynamics}\label{appA}

In the literature, the quadratic Poincar\'e gravity theories are often formulated in terms of the standard tensor objects which are not decomposed into irreducible parts. As usual, we introduce the {\it Ricci tensor} and the {\it co-Ricci tensor} as 
\begin{equation}\label{Ric}
R_{ij} := R_{kij}{}^k,\qquad \overline{R}{}^{ij} := {\frac 12}\,R_{klm}{}^i\,\eta^{klmj},
\end{equation}
from which the curvature scalar and pseudoscalar arise naturally as the traces
\begin{equation}\label{scalars}
R = g^{ij}R_{ij} = R_{ij}{}^{ji},\qquad \overline{R} =
g_{ij}\overline{R}{}^{ij} = {\frac 12}\,R_{ijkl}\,\eta^{ijkl}.
\end{equation}
We consider the general quadratic model with the Yag-Mills type Lagrangian that contains all possible quadratic invariants of the torsion and the curvature:
\begin{align}
L &= -\,{\frac 1{2\kappa c}}\Bigl\{a_0 R + \overline{a}{}_0\overline{R} + 2\lambda_0\nonumber\\ 
& \hspace{12mm} +\,a_1\,T_{kl}{}^i\,T^{kl}{}_i + a_2\,T_i\,T^i + a_3\,T_{kl}{}^i\,T_i{}^{kl}\nonumber\\
& \hspace{12mm} +\,\overline{a}_1\,\eta^{klmn}\,T_{kli}\,T_{mn}{}^i + \overline{a}_2\,\eta^{klmn}
\,T_{klm}\,T_n\nonumber\\
&\hspace{5mm} +\,{\ell_\rho^2}\Big(b_1\,R_{ijkl}R^{ijkl} 
+ b_2\,R_{ijkl}R^{klij} + b_3\,R_{ijkl}R^{ikjl} \nonumber\\
&\hspace{12mm} +\,b_4\,R_{ij}R^{ij} + b_5\,R_{ij}R^{ji} + b_6\,R^2\nonumber\\
&\hspace{12mm} +\,\overline{b}_1\,\eta^{klmn}\,R_{klij}R_{mn}{}^{ij}
+ \overline{b}_2\,\eta^{klmn}\,R_{kl}R_{mn}\nonumber\\
&\hspace{12mm} +\,\overline{b}_3\,\eta^{klmn}\,R_{klm}{}^i\,R_{ni}
+ \overline{b}_4\,\eta^{klmn}\,R_{klmn}\,R\Big)\Bigr\}.\label{LRT}
\end{align}
Here $\kappa = {\frac {8\pi G}{c^4}}$ is Einstein's gravitational constant with the dimension of $[\kappa c] = $s\,kg$^{-1}$. $G = 6.67\times 10^{-11}$ m$^3$\,kg$^{-1}$\,s$^{-2}$ is Newton's gravitational constant. The speed of light $c = 2.9\times 10^8$ m/s. For completeness, we include the cosmological term $\lambda_0$.

Besides the linear ``Hilbert type'' part characterized by $a_0$ and $\overline{a}_0$, the Lagrangian (\ref{LRT}) contains several additional coupling constants which fix the structure of the ``Yang-Mills type'' part: $a_1, a_2, a_3$, $\overline{a}_1, \overline{a}_2$, $b_1, \cdots, b_6$, $\overline{b}_1, \cdots, \overline{b}_4$, and $\ell_\rho^2$. The coupling constants $a_I$, $\overline{a}_I$, $b_I$ and $\overline{b}_I$ are dimensionless, whereas the dimension $[{\ell_\rho^2}]=\,$[area] so that $[{\ell_\rho^2}/{\kappa c}] = [\hbar]$.

The analysis of the particle spectrum for the quadratic model (\ref{LRT}) reveals that the dynamics of gravitational modes in different $J^P$ (spin$^{parity}$) sectors is determined by the following combinations of the coupling constants: $2^\pm$ sector 
\begin{eqnarray}
\Lambda_1  = 4(b_1 + b_2) + 2b_3 + b_4 + b_5,\qquad \Lambda_2 = 4b_1 + b_3,\label{L12}
\end{eqnarray}
$0^\pm$ sector 
\begin{eqnarray}
\Lambda_3 = 4(b_1 + b_2) + 2b_3 + 4(b_4 + b_5) + 12b_6,\qquad \Lambda_4 = 4b_1 - 2b_3,\label{L34}
\end{eqnarray}
and $1^\pm$ sector
\begin{eqnarray}
\Lambda_5 = 4(b_1 - b_2) + b_4 - b_5, &\qquad& \Lambda_6 = 4b_1 + b_3 + 2b_4,\label{L56}\\
\overline{\Lambda}_5 = 4(\overline{b}_2 - 2\overline{b}_3), &\qquad&
\overline{\Lambda}_6 = -\,4(\overline{b}_2 + 3\overline{b}_4),\label{L56o}
\end{eqnarray}
whereas the mass terms are specified by
\begin{eqnarray}
\mu_1 &=& -\,a_0 + a_1 - a_2,\qquad \mu_2 = -\,2a_0 + {\frac {2a_1 + a_2 + 3a_3}{4}},\label{mu1}\\
\mu_3 &=& -\,a_0 - {\frac {2a_1 + a_2}{4}},\qquad \overline{\mu}_1 = {\frac 83}\left(
4\overline{a}_1 + 3\overline{a}_2 - 2\overline{a}_0\right).\label{mu3}
\end{eqnarray}


\end{document}